\documentclass[journal]{IEEEtran}
\usepackage[cmex10]{amsmath}
\usepackage{exscale,times}
\usepackage[T1]{fontenc}
\usepackage{amssymb}
\usepackage{color}
\usepackage{cite}
\usepackage{graphicx}
\usepackage{epsfig}
\usepackage{multirow}
\usepackage{float}
\usepackage{caption}
\usepackage{subcaption}
\usepackage{url}
\usepackage{dblfloatfix} 
\usepackage{makecell}
\begin{document}

\title{sEMG-Based Natural Control Interface for\\ a Variable Stiffness Transradial Hand Prosthesis}

\author{Elif~Hocaoglu,~\IEEEmembership{Member,~IEEE}
        and~Volkan~Patoglu,~\IEEEmembership{Member,~IEEE}
\thanks{E. Hocaoglu and V. Patoglu are with the Faculty of Engineering and Natural Sciences of Sabanc\i\ University, \.Istanbul, Turkey.
       {\tt\footnotesize \{elifhocaoglu,vpatoglu\}@sabanciuniv.edu}}%
\thanks{Manuscript received October 25, 2019; revised November XX, 2019.}}

\markboth{IEEE/ASME TRANSACTIONS ON MECHATRONICS, VOL. XX, NO. X, OCTOBER 2019}%
{Shell \MakeLowercase{\textit{et al.}}: Bare Demo of IEEEtran.cls for Journals}

\maketitle

\begin{abstract}
We propose, implement and evaluate a natural human-machine control interface for a variable stiffness transradial hand prosthesis that achieves tele-impedance control through surface electromyography (sEMG) signals. This interface, together with variable stiffness actuation (VSA), enables an amputee to modulate the impedance of the prosthetic limb to properly match the requirements of a task, while performing activities of daily living. Both the desired position and stiffness references are estimated through sEMG signals and used to control the VSA hand prosthesis. In particular, regulation of hand impedance is managed through the impedance measurements of the intact upper arm; this control takes place naturally and automatically as the amputee interacts with the environment, while position of the hand prosthesis is regulated intentionally by the amputee through the estimated position of the shoulder. The proposed approach is advantageous, since the impedance regulation takes place naturally without requiring amputees' attention and diminishing their functional capability. Consequently, the proposed interface is easy to use, does not require long training periods or interferes with the control of intact body segments. The performance of the approach is evaluated through human subject experiments where adequate estimation of references and independent control of position and stiffness are demonstrated.
\end{abstract}


%
\IEEEpeerreviewmaketitle
\vspace{-\baselineskip}
\section{Introduction}
\label{Sec:Introduction}

\IEEEPARstart{A}{ccording} to the World Health Organization, there are about 40 million amputees living in the developing countries~\cite{Marino15}, and this number is expected to rise in the future~\cite{Graham2007}. The report published by the National Center for Health Statistics states that every year about 50000 people are amputated in the US and the ratio of upper extremity to lower extremity amputation in this population is 1 to 4~\cite{Lusardi13}. Many prosthetic devices have been proposed to raise the life standards of amputees by helping them restore their functional abilities, enabling them to perform daily chores and return back to their work~\cite{Millstein}.

Despite many potential benefits, a substantial percentage of people with upper-limb amputation prefer not to wear prostheses. In the literature, the mean rejection rates for the use of electric and body-powered prostheses are reported for pediatric population as 35\% and 45\%, and for the adult population as 23\% and 26\%, respectively~\cite{Biddiss}. Some of the reasons behind the low acceptance rate of body-powered hands are reported as slow movement, heavy weight, inadequate grip force, limited functionality, inconvenience of harnessing, unnatural use, and discomfort~\cite{Biddiss}. Myoelectrically controlled powered prostheses provide some advantages over the body-powered ones in terms of providing higher grip strength with less fatigue, more functionality, and less harnessing. Notwithstanding such benefits, the consensus is that they give rise to pain, have high cost, are heavy and inconvenient due to the need to regularly charge the battery and provide periodic maintenance~\cite{Biddiss,McFarland}. All these factors negatively influence their acceptance.

Many research groups have investigated means to close the acceptance gap by orienting their studies to increase the dexterity and functionality of prosthetic hand devices. In both academic studies~\cite{Wang,Naik,Riillo} and commercial applications~\cite{ilimb,Ottobockhand,bebionic,mchand}, the most common means to  control dexterous hand prosthesis is based on classifying sEMG signals recorded from different muscle groups and assigning a grip pattern to each class. Recently, some studies have also integrated different sources of data, such as MMG~\cite{Kurzynski}, NIRS~\cite{Guo}, IMU~\cite{Kyranou}, to improve the classification performance of multi-functional hand prostheses. Although such studies are aimed to make the amputees' life easier by enabling hand prostheses to have more functions, these devices demand long-training periods~\cite{Herle} stemming from their non-intuitive control interface, and have not been shown to provide a viable solution for the high abandonment rate of prosthetic devices~\cite{Atkins}. %

To enable natural dexterity and an intuitive control interface for prosthetic hand devices, one of the prominent features of human neuromuscular system specialized to be competent at realizing various physical activities may provide a solution. In particular, most of the daily activities requiring physical interactions with human hands are successfully performed thanks to the unrivalled capability of human adaptation. Such ability orginates from predicting the type of the interaction and regulating the impedance of the limb based on the activity~\cite{Perreault,Popescu,Franklin2003,Milner2003,Franklin2004,Kawato2003}. The impedance regulation of limbs is realized through the modulation of the contraction levels of antagonistic muscle pairs, as well as reflexive reactions that contribute to neuromotor control  to assist the stability of human-object interaction. All these abilities enable humans to actively and naturally perform activities of daily living (ADL). For instance, during tasks that require high precision (such as writing or drilling a hole), humans raise the stiffness of their arm to guarantee the precise positioning against perturbations, while during interactions with soft/fragile objects, humans regulate their limbs to become more compliant in order to prevent damage to the object~\cite{hogan2002}.

The impedance modulation ability of humans has become inspiring in robotics. Along these lines, several studies on prosthetic devices have been conducted to imitate the stiffness regulation feature of humans, while physically interacting with their environment~\cite{hoganpart1,hoganpart2,twente}. Moreover, systematic human subject experiments have provided evidence that task-dependent impedance regulation improves the human performance while using a virtual arm prosthesis~\cite{Okamura2011,okamura2012,okamura2013}. Recently, authors have proposed a variable stiffness transradial hand prosthesis~\cite{hocaoglue12,Part2}. Variable stiffness actuation (VSA) of this prosthesis is based on antagonistically arranged tendons coupled to nonlinear springs driven through a Bowden cable based power transmission.  This variable stiffness transradial hand prosthesis features tendon driven underactuated compliant fingers that enable natural adaption of the hand shape to wrap around a wide variety of object geometries and modulation of hand's impedance to perform various tasks with high dexterity. Unlike in the control based impedance modulation, VSA based prosthesis possesses high energy efficiently, since its actuators are not in use at all times to maintain a desired stiffness level. Furthermore, since the resulting stiffness of VSA is an inherent physical property of the device, it is valid over the whole frequency spectrum, including the frequencies over the controllable bandwidth of the actuators.

sEMG signals have been commonly utilized for motion (position or velocity) control of prosthetic devices~\cite{myohand1,myohand2,myohand3,Kim2019}. In particular, commercial hand prostheses, such as i-Limb Ultra~\cite{ilimb}, Sensor Hand~\cite{Ottobockhand}, Bebionic Hand~\cite{bebionic}, and Motion Control Hand~\cite{mchand} all rely on motion controllers executed proportional to sEMG signals measured from an amputee.  These devices require the amputee intentionally modulate sEMG levels of several distinct muscle groups, such that the motion 
of the prosthetic/robotic device can be controlled.

\begin{figure*}[t]
\centering
\resizebox{1.775\columnwidth}{!}{\includegraphics{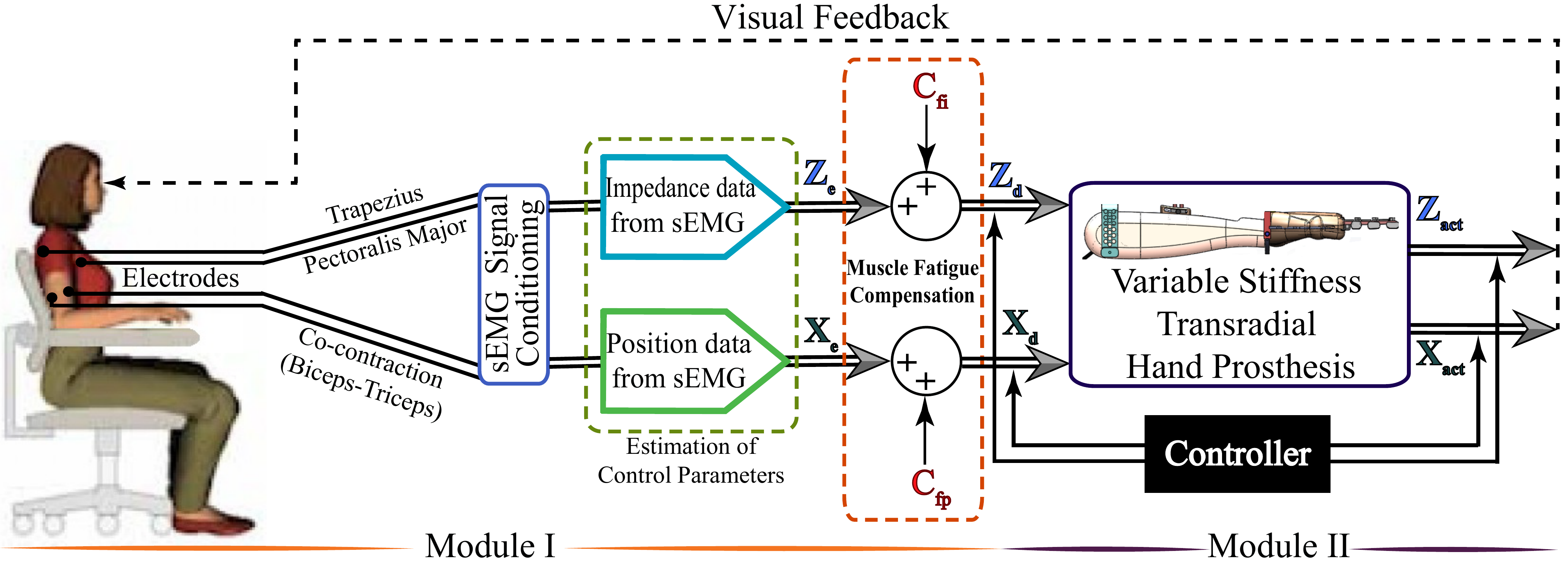}}
\vspace*{-0.25\baselineskip}
\caption{The control interface of the variable stiffness transradial hand prosthesis: In the first module, raw sEMG signals are measured from the upper arm and muscle groups placed under the chest and shoulder, sEMG signals are conditioned by a series of filters, and the position and stiffness references are estimated. The second module implements the position and stiffness control of the variable stiffness transradial hand prosthesis to follow the references estimated in the first module.}
\vspace*{-0.25\baselineskip}
\label{control}
\end{figure*}

In this study, we propose, implement and evaluate a natural human-machine interface for a variable stiffness transradial hand prosthesis to achieve tele-impedance control through sEMG signals. The mechatronic design of the transradial hand prosthesis, presented in~\cite{Part1}, employs a VSA based on antagonistic actuation principle with quadratic springs and  enables amputees to regulate the stiffness and position of the hand prosthesis independently. For the tele-impedance control of the variable stiffness transradial hand prosthesis, we benefit from sEMG signals generated during the muscular activity captured by biopotential electrodes, by means of which amputees can naturally be a part of the control architecture.

Our human machine interface is based on using four channels of sEMG signals responsible for controlling position and impedance of the variable stiffness transradial hand prosthesis. In particular, as commonly done in the literature, the motion control of the hand prosthesis is regulated through intentional muscular activities generated at chest and shoulder mapped to the opening/closing of the fingers. However, in contrast to other interfaces, the stiffness of the prosthesis is regulated automatically based on the estimated stiffness of the \emph{intact} muscle groups of the upper arm.  As a result, while the proposed human machine interface requires amputee to intentionally control the position of the VSA prosthesis, the stiffness regulation takes place automatically based on the instantaneous stiffness of the intact portion of the limb. Such an approach is advantageous, since the impedance regulation takes place effortlessly  from task to task or during execution of a single task without requiring amputees' attention and diminishing their functional capability. Consequently, such an interface is easy to use, does not require long training periods, and does not interfere with the control of intact body segments. Furthermore, it has been pointed out in the literature that energetic interactions with the environment  influence the determination of the impedance by the intact neuromuscular system~\cite{Franklin2004}. Hence, regulating the prosthesis to mimic the impedance of an intact portion of the limb promises to be a more plausible control strategy than requiring the amputee to determine and control the proper impedance using dysfunctional muscles that lack such physical feedback, since these muscles are not physically coupled to the environment.

A preliminary study regarding tele-impedance control of variable stiffness transradial hand prosthesis has been presented in~\cite{hocaoglue12}. This study significantly extends~\cite{hocaoglue12} by employing different muscles groups that are better suited for the task, experimentally verifying the correlated impedance modulation of the antagonistic muscle groups at the forearm and the upper arm, incorporating a compensation strategy for fatigue induced by  prolonged and repeated co-contraction of muscles, and providing an experimental evaluation of the natural human-machine interface with a variable stiffness transradial hand prosthesis under different tasks, while interacting with fragile, soft, rigid, complex shaped objects. To the best of authors' knowledge, this study, along with~\cite{hocaoglue12}, present one of the first human-machine control interfaces for a VSA hand prosthesis. Furthermore, the human subject experiments presented in this study complements the ones in the literature~\cite{Okamura2011,okamura2012,hocaoglue12,okamura2013}, as physical interactions with the environment are enabled in this study.

The rest of the paper is organized as follows: Section~II introduces sEMG-based tele-impedance control of the variable stiffness transradial hand prosthesis. Sections~III and~IV  present the construction of control references through sEMG based stiffness and position estimations, respectively. Section~V explains the independent control of the position and stiffness for a antagonist VSA. Section~VI details the compensation strategy used against muscle fatigue. Section~VII presents a set of experiments to verify the hypothesis that the stiffness modulation of the upper arm and the forearm are correlated.  Section~VIII experimentally verifies the independent control of the position and stiffness through the sEMG-based tele-impedance control of a VSA transradial hand prosthesis and provide evidence that the natural human-machine interface is an effective strategy in control. Section~IX demonstrates the grasping performance of the tele-impedance controlled VSA transradial hand prosthesis on manipulation of various objects. Finally, Section~X concludes the study and discusses future work.

\section{sEMG Based Tele-Impedance Control of a Variable Stiffness Transradial Hand Prosthesis}
\label{Sec:TeleImpedanceControl}

sEMG based tele-impedance controller is developed to control a VSA transradial hand prosthesis~\cite{Part1}. The transradial hand prosthesis features tendon driven underactuated compliant fingers that naturally adapt the hand shape to wrap around a wide variety of object geometries. Antagonistically arranged tendons of the prosthesis enable the modulation of the stiffness of the fingers, as well as control of their position. Adaptation of the mechanical impedance of prosthesis based on changing physical conditions enables the amputee to perform various tasks with high dexterity.

Figure~\ref{control} presents an overview of the tele-impedance control architecture. The proposed control architecture consists of two modules. The first module handles the measurement of sEMG signals, their conditioning, and the estimation of reference values for the hand position and stiffness. The second module includes a closed loop controller that ensures that the position and the stiffness of the VSA prosthetic hand match these reference values. Throughout the control, visual feedback and physical coupling provide information for the amputees to adapt their sEMG signals to match the task requirements.

In particular, given that transradial upper extremity amputees lack the muscle groups responsible for hand and forearm motions, sEMG signals for the position control of the hand prosthesis are measured from the chest and the shoulder, while sEMG signals measured from the intact muscle pairs on the upper arm are used for the impedance control. Estimation of the hand position and stiffness from sEMG signal involves modeling of hand motion/stiffness based on sEMG signals, empirical determination of model parameters for use in real time control, and incorporation of fatigue compensation.

\section{Stiffness Estimation through sEMG Signals}
\label{Sec:ImpedanceEstimation}

Muscle groups play a crucial role in the human body in terms of both the torque and impedance (stiffness and damping) modulation of a joint to properly interact with different environmental conditions. Particularly, impedance matching to the varying environment dynamics is carried out by means of the prominent features of muscles, such as regulation of co-contraction levels and reflex gains. The mechanical impedance of joints is an important parameter in the control of limbs under both static and dynamic conditions.

In the literature, many researchers have addressed the characterization of joint stiffness by focusing on multi-joint arm movements~\cite{Gomi05041996,Burdet2000,Burdet2001}. These studies are mainly focused on point-to-point reaching movements of subjects  under perturbations and disturbance forces. The stiffness of the arm is estimated based on the relation between the deviations of the trajectories with respect to the undisturbed trajectories and the applied perturbation forces. Such methods are not viable for real-time applications, such as use with prosthetic limbs, as they require coupling users to a computer controlled manipulator. Index of muscle co-contraction around the joint (IMCJ)~\cite{shortandlong} approach is based on sEMG signals and provides a stiffness estimation technique that is feasible for a real-time use. In this approach, the stiffness estimation is realized through estimation of the co-contraction levels of antagonistic muscle groups. In the literature, IMCJ method has been employed to reveal the mechanical characteristics of the musculoskeletal system~\cite{Gomi1998,Hunter_Kearney_1982,Basmajian}.

IMCJ describes the working principle of antagonistic muscle groups around a joint through rectified sEMG signals and utilizes Eqns.~(\ref{stiffness1})-(\ref{stiffness2}) for stiffness estimation of the joint~\cite{shortandlong}.
\begin{eqnarray}
\label{stiffness1}
\tau\!\!&=&\!\!\sum_{i=1}^k \big[\kappa_{i}. agon(sEMG_{i})-\lambda_{i}. anta(sEMG_{i})\big]\\
\label{stiffness2}
S\!\!&=&\!\!\sum_{i=1}^k \big[|\kappa_{i}|. agon(sEMG_{i})+|\lambda_{i}|. anta(sEMG_{i})\big]
\end{eqnarray}
Here, $i$ is the index that labels each muscle group, $\tau$ symbolizes the joint torque of the limb, while $agon(sEMG)$ and $anta(sEMG)$ denote the normalized muscular activity of the agonist and antagonist muscles, respectively. Symbols $\kappa$ and $\lambda$ capture the moment arms in charge of converting muscle activity to muscle tension responsible for generating muscle torque. The relation between the muscle torque and the muscle impedance~\cite{Gomi1998,Kuechle1997,Murray1995} is mapped to the correlation between the joint torque and the joint impedance~\cite{Gomi1998,Hunter_Kearney_1982}, leading to the joint stiffness estimates $S$ via Eqn.~(\ref{stiffness2}), where  $\kappa$ and $\lambda$ are estimated according to Eqn.~(\ref{stiffness1}).

In this study, Eqns.~(\ref{stiffness1})-(\ref{stiffness2}) were used to  estimate the joint stiffness through a series of experiments as follows. Eight healthy volunteers (2 females, 6 males), who were students of Sabanc{\i} University  participated in the experiments. Participants had no prior experience with the experimental setup. The participants did not report any sensory or motor impairment. The participants signed informed consent forms approved by the University Research Ethics Council of Sabanc{\i} University.

\begin{figure}[h!]
\centering
\resizebox{2.35in}{!}{\includegraphics{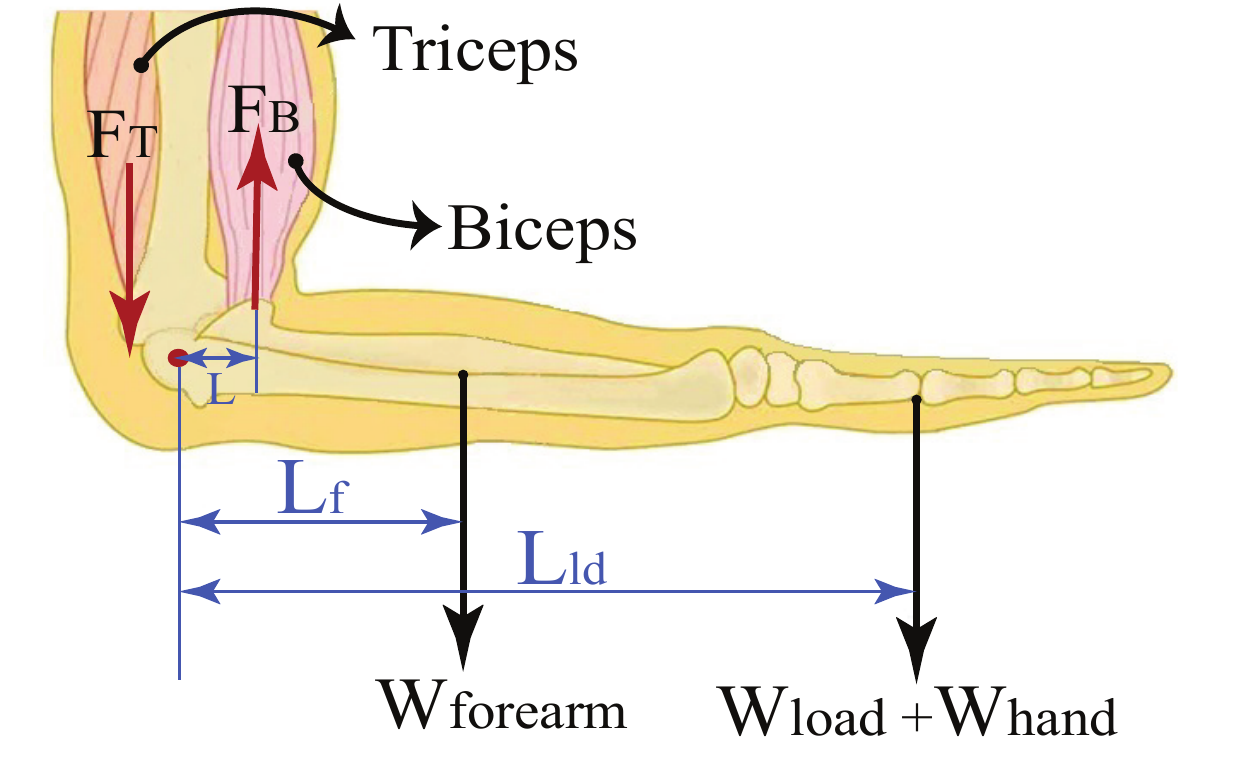}}
\vspace*{-0.5\baselineskip}
\caption{Biomechanical model with the pivot at the elbow joint, and the elbow positioned at 90$^{\circ}$}
\label{biomechanic_models1}
\end{figure}

The experimental task was to grasp a dumbbell while positioning the elbow at 90$^{\circ}$, as shown in Figure~\ref{biomechanic_models1}. In particular, the forearm was configured horizontally, while the upper arm was kept perpendicular to the forearm with the palm was facing down. To maintain this configuration, the antagonistic muscle groups placed on upper arm were isometrically contracted not to change the palm configuration and to exert appropriate forces to keep the joint angle at the desired value.

Participants started by lifting their forearm when their hand was free, and then the load was gradually increased using dumbbells of 0.5kg, 1kg, 1.28kg, 2.26kg, 2.76kg, and 3.76kg, respectively. Each condition was tested for 20 trials, where each trial lasted 20 seconds, on average.

The net torque applied at the elbow joint is calculated using the weight of the load $W_{load}$ and the weight of the forearm $W_{forearm}$ together with the moment arm corresponding to the load $L_{ld}$ and  the center of gravity of the forearm  $L_{f}$ with respect to the elbow joint.

\begin{figure}[b!]
\centering
\resizebox{.95\columnwidth}{!}{\includegraphics{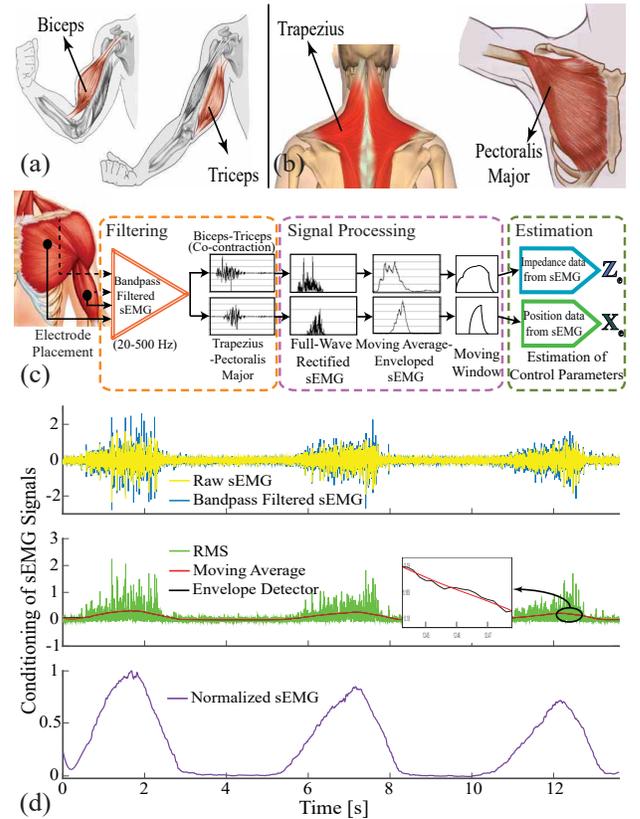}}
\vspace*{-.5\baselineskip}
\caption{(a) Biceps and triceps muscles are responsible for the stiffness modulation. b) Trapezius and pectoralis major muscles are employed for position regulation. (c) sEMG signal flow: Raw sEMG signals (yellow) are bandpass filtered (blue) and full wave rectified. Then, these signals are averaged  using 0.5 second moving window and undesired ripples are omitted by means of envelope detection. (d)  On the top graph, the raw sEMG data filtered against the inherent and environmental noises, and artifacts are represented with the blue signal. The second graph depicts the rectified (green), moving averaged (red), enveloped (black) sEMG signal. The bottom  figure shows the normalized sEMG signal.}
\label{signalconditioning_example_musclegroups}
\end{figure}

The antagonistic muscle pairs, biceps and triceps, responsible for generating the sEMG signals for the stiffness estimation are shown in Figure~\ref{signalconditioning_example_musclegroups}(a). sEMG signals were measured by means of  surface electrodes of an sEMG signal acquisition device with the sampling rate of 1~kHz. Raw sEMG signals were collected during the trials and conditioned by means of a full-wave rectifier, a moving average filter and an envelope detector. During the analysis, the first 500 samples of each trials were omitted from the experimental data to exclude signal outliers owing to initialization and motion artifacts.

The signal conditioning process is illustrated in Figure~\ref{signalconditioning_example_musclegroups}(c), while a sample signal extracted from a real-time experiment is presented in Figure~\ref{signalconditioning_example_musclegroups}(d). In particular, raw sEMG data was filtered against inherent and environmental noises and motion artifacts utilizing a Butterworth band-pass filter with a frequency range of 20-500 Hz. After increasing the signal-to-noise ratio of the sEMG signal, full wave rectified signal was obtained. To filter out the ripples of the rectified signal, an envelope detector and a moving window over a period of 0.5 seconds were employed. Such conditioning methods play crucial role in revealing the relation between joint torque and sEMG signals. Finally, normalization of sEMG signals was carried out using the maximum voluntary contraction (MVC) of the participants.

\begin{table}[t!]
\caption{Estimated parameters of the stiffness model}
\vspace*{-.5\baselineskip}
\label{imp_param}
\begin{center}
\begin{tabular}{c | c |c | c  }
$\kappa$ & $\lambda$ & R$^2$ & RMSE \\
\hline \hline
1.8612 & 1 & 0.9938 & 0.02941
\end{tabular}
\end{center}\vspace*{-1.5\baselineskip}
\end{table}

The parameters in Eqn.~(\ref{stiffness1}) were estimated using multiple linear regression by means of recorded data streams of $agon(sEMG)$, $anta(sEMG)$ and $\tau$. For this linear model, the regression coefficients were obtained with 95\% confidence bounds. The estimations for a subject are presented in Table~\ref{imp_param}. The quality of the estimation for all subjects was evaluated to be high, with $R^2>$0.99 and RMSE < 0.03.

\section{Position Estimation through sEMG Signals}
\label{Sec:PositionEstimation}

Position of the underactuated variable stiffness prosthetic hand is controlled intentionally under visual feedback. As the prosthesis is underactuated, the exact positions of the fingers depend on the interaction between the prosthesis and the environment, as well as the position controller tracking the reference signal generated by the amputee. In this application, the precise estimation of position reference is not of critical importance, since the amputee can adjust the position of the prosthetic hand based on visual feedback.

The position of the transradial prosthetic hand is controlled through a direct proportional relation between the intensity of sEMG signals with the desired joint angle. In order achieve independent and simultaneous position and stiffness control, the overlap of sEMG signals corresponding to the stiffness reference with sEMG signals corresponding to the position reference has to be avoided. All muscle groups on the arm take part in the isometric contraction. Since sEMG signals measured from the upper arm is used to the estimate stiffness reference, to avoid any overlap, pectoralis major and trapezius muscles placed in the chest and shoulder, shown in Figure~\ref{signalconditioning_example_musclegroups}(b), are preferred for the position control of hand prosthesis. This selection ensures independent location of muscle pairs responsible for stiffness modulation and position control, such that their activities do not directly affect each other.

Another design parameter while constructing the position control references is the MVC percentage that is used for normalization. Instead of mapping 100\% MVC to fully close/open the hand, a lower MVC can be set to decrease the muscle fatigue to a great extent. In our study, the MVC level is selected as 70\%, such that the position reference for the actuation of VSAs is calculated using the following normalized sEMG signal
\begin{equation}
\label{isometric}
sEMG_{normpos}=\frac{sEMG_{position}-sEMG_{bias}}{sEMG_{ \%70MVC}}
\end{equation}

\noindent where $sEMG_{normpos}$ denotes the normalized sEMG signals corresponding to position reference, $sEMG_{position}$ represents the conditioned sEMG signals measured from pectoralis major and trapezius muscles, $sEMG_{\%70MVC}$ is 70\% MVC of the responsible muscles, and $sEMG_{bias}$ is the bias on the signal.

Another undesirable condition is the contamination of sEMG signals generated by pectoralis major responsible for opening of the hand by electrocardiography (ECG) signals. ECG crosstalk effect is prevented from sEMG signals by avoiding the electrode placement in the contamination zone and by adding extra bias term to the sEMG signals until ECG signal effect is suppressed.

\section{Compensation against Muscle Fatigue}
\label{Sec:FatigueEffect}

Muscle fatigue can be defined as a decline in the muscle strength to generate force, that is, a decrease in the sEMG amplitude as a result of reduction in active muscle fibers during ceaseless muscle activity~\cite{mus_ftg4}. The reason of muscle fatigue encompasses the metabolic, structural and energetic alternations in muscles owing to insufficient oxygen level, inadequate blood circulation responsible for supplying nutritive substances, and also decrease in the efficiency of the nervous system~\cite{mus_ftg2}.

\begin{figure}[b]
\centering \vspace*{-.5\baselineskip}
\resizebox{\columnwidth}{!}{\includegraphics{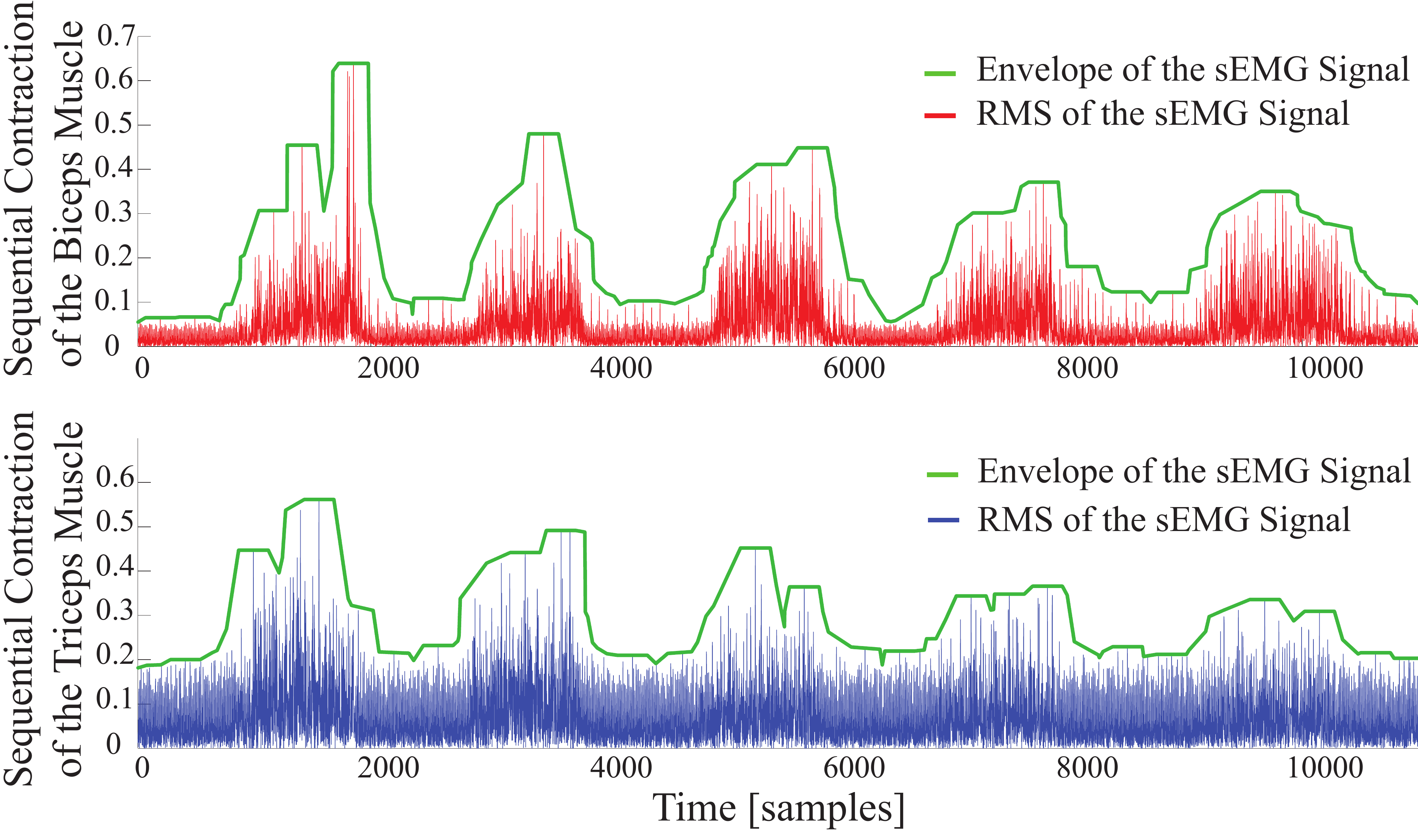}}
\vspace*{-1.5\baselineskip}
\caption{sEMG signal features capturing the average fatigue characteristics of biceps and triceps muscles}
\label{fatigue_example}
\end{figure}

Myoelectric signals collected on the surface of the skin can be used for real-time monitoring of muscle fatigue~\cite{deluca84}.  This method is commonly preferred since it can provide uninterrupted data recordings related to muscle fatigue with a non-invasive technique, even though this method has certain disadvantages, such as the difficulties associated with exact positioning of surface electrodes on desired muscles and undesired cross-talk of the myoelectric signals with the neighboring muscles.  A large number of studies have been performed to establish a signal-based quantitative criteria to characterize muscle fatigue under static and dynamic tasks. Along these lines, numerous classical and modern signal processing methods have been established for sEMG-based muscle fatigue evaluation~\cite{mus_ftg1}.

In this study, we rely on a time-domain root-mean-square (RMS) feature of sEMG signals to compensate for the fatigue effect~\cite{mus_ftg3}. In particular, during the use of the prosthetic device, the muscle performance decreases as a function of use time; as muscle fatigue increases, the sEMG-based stiffness reference estimates deteriorate. RMS feature based fatigue compensation estimates the decrease in sEMG signal power as a function of use time and introduces a compensation factor to counteract this fatigue.

Muscle fatigue compensation is activated in the control loop when a threshold is exceeded. A moving average window of 2000 samples runs to check the presence of the consecutive contractions, by comparing the average level of enveloped sEMG signal under the moving window with the threshold. Threshold commissioned for activation of the fatigue compensation is empirically determined as about $20\%$ MVC and varies slightly among volunteers. Figure~\ref{fatigue_example} illustrates muscle fatigue captured by sEMG signals, when a volunteer repeatedly co-contracts her muscles within 5 seconds. In the figure, the green line presents the envelope of RMS of sEMG signals and the decrease of signal power can be observed. 

To estimate the fatigue characteristics from sEMG signals, an experiment is conducted where a volunteer is requested to realize sustained isometric contractions periodically. The experiment includes 10 sessions with each session including 5 trials and lasting for 30 seconds.

The fatigue behavior of the individual is extracted from the sEMG data through three sequential signal conditioning stages. Firstly, the raw sEMG signals are band-pass filtered with a frequency band between 20-500~Hz to remove undesired signals due to electronic noise, motion artifacts, ECG cross-talk, and power-line interference. Secondly, filtered sEMG signal is normalized with the MVC of the volunteer. Finally, RMS of sEMG signal is calculated.

\begin{figure}[htb]
\centering
\resizebox{.9\columnwidth}{!}{\includegraphics{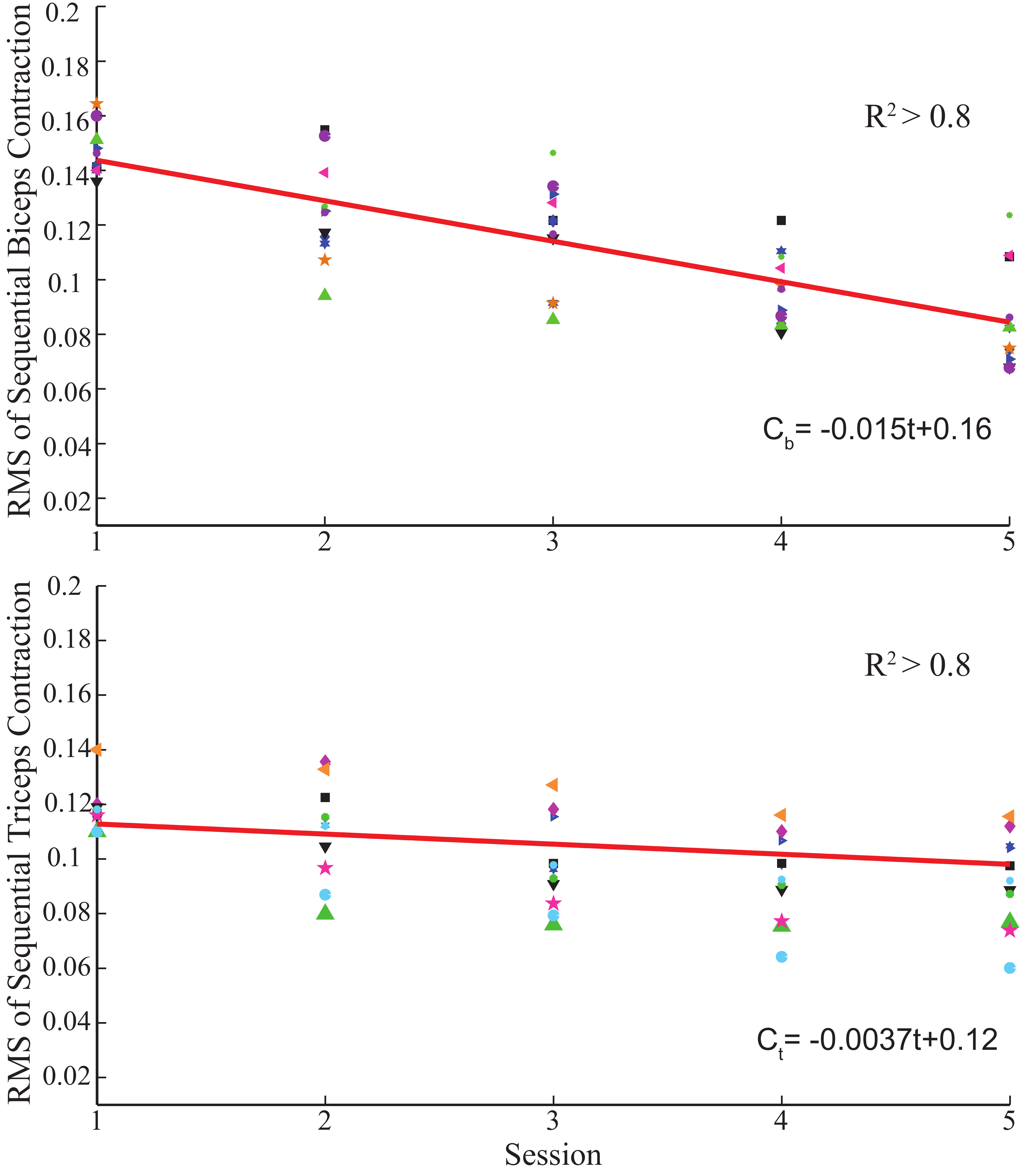}}
\vspace*{-.75\baselineskip}
\caption{Linear fits capturing the average fatigue characteristics of biceps and triceps muscles}
\label{fatigue}
\vspace*{-1\baselineskip}
\end{figure}

Figure~\ref{fatigue} presents sample results characterizing the fatigue observed on biceps and triceps muscles as a function of the use time. Linear fits, as presented in Figure~\ref{fatigue}, are sufficient to capture the time dependent fatigue characteristics embedded in this data set, as evidenced by good quality  of  curve fits ($R^2>0.8$). Once these linear estimates are at hand, they can be incorporated in the stiffness reference estimation as a feed-forward compensation term denoted by $C_{fi}$ in Figure~\ref{control}. Unlike the impedance modulation, position control typically does not require sequential contractions; hence, the muscle fatigue is neglected during position regulation, that is, no feed-forward compensation is performed for the position control by setting $C_{fp}=0$ in Figure~\ref{control}.

\section{Position and Stiffness Regulation with Antagonist VSA}

\begin{figure}[t]
\centering
\resizebox{0.6\columnwidth}{!}{\includegraphics{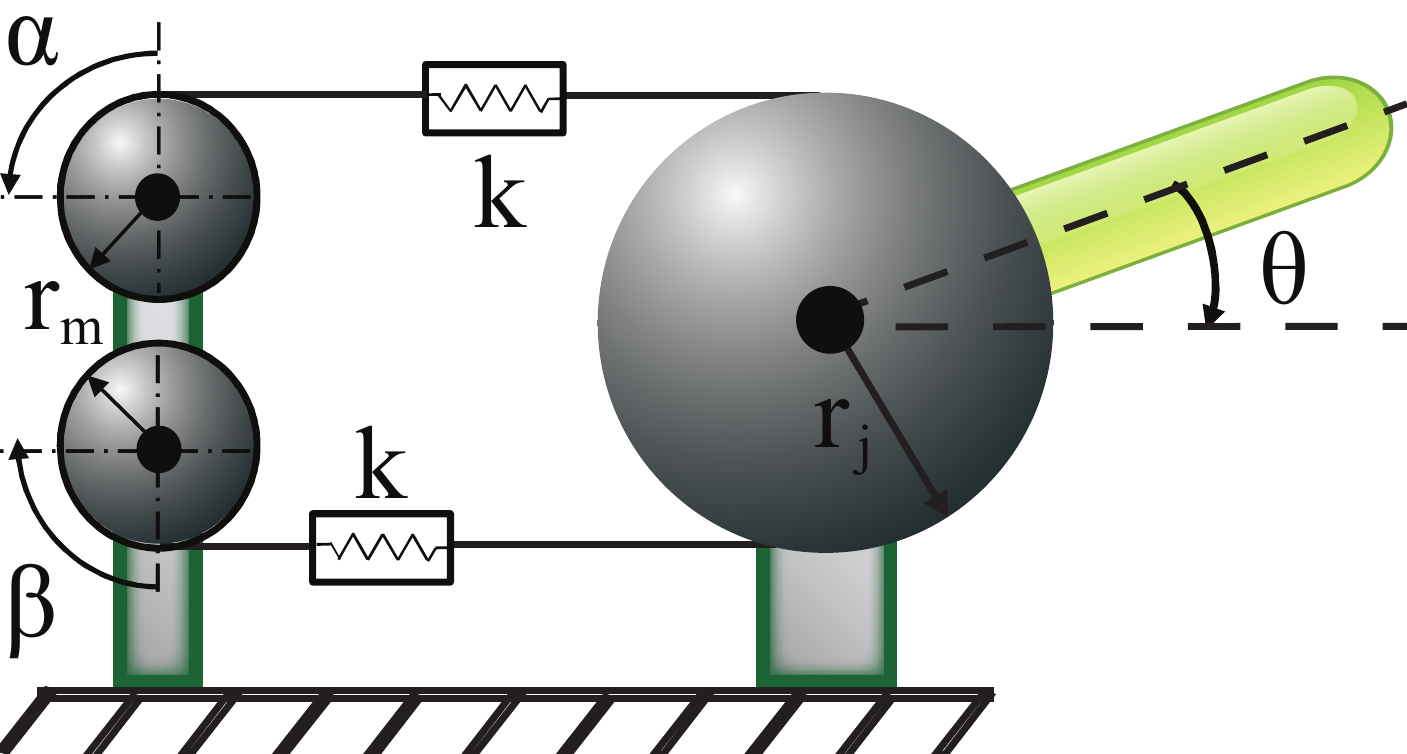}}
\vspace*{-0.2\baselineskip}
\caption{Schematic model of a antagonistically driven VSA}
\label{schematic_exp_vsa_red}
\vspace*{-.5\baselineskip}
\end{figure}

Given the sEMG based position and stiffness reference estimation and fatigue compensation processes, the second module of the interface is a controller that ensures tracking of these references by the VSA prosthetic hand. In particular, the position and stiffness of the VSA are controlled through position control of Bowden cables driven by two geared DC motors. Figure~\ref{schematic_exp_vsa_red} presents a schematic representation of the VSA, where $\alpha$ and $\beta$ denote angular position of DC motors, while $S$ and $\theta$ represent the joint stiffness and angle, respectively.

Under quasi-static conditions~\cite{Migliore2007,EnglishRussell}, the angular position of DC motors $\alpha$ and $\beta$ for a given reference position $\theta_r$ and  stiffness $S_r$ can be calculated as
\begin{eqnarray}
\alpha=(S_r-2 b r_j^2)/4 a r_m r_j^2+(r_j/r_m)((\tau_{load}/S_r)-\theta_r) \label{thetastiffeqn1}\\
\beta=(S_r-2 b r_j^2)/4 a r_m r_j^2-(r_j/r_m)((\tau_{load}/S_r)-\theta_r) \label{thetastiffeqn2}
\end{eqnarray}
\noindent where $r_{m}$ represents the radius of the pulleys attached to the geared DC motors, $r_j$ is the radius of the drive pulley, $a$ and $b$ are the parameters that characterize the expanding contour cam as deatiled in~\cite{Part1}, while the external torque applied to VSA is denoted by $\tau_{load}$.
When control references belonging to joint position and stiffness are estimated through sEMG signals, desired motor positions are computed according to Eqns.~(\ref{thetastiffeqn1})-(\ref{thetastiffeqn2}) with $\tau_{load}=0$  and motors are motion controlled to these values under real-time control. 

\section{Verification of Correlated Stiffness Adaptation of Antagonistic Muscle Pairs} 
\label{Sec:SimultaneousImpedance}

The stiffness of the prosthesis is regulated automatically based on the estimated stiffness of the \emph{intact} muscle groups of the upper arm, such that impedance regulation takes place automatically and naturally  from task to task or during execution of a single task without requiring amputees' attention and diminishing their functional capability. This control strategy, in which the prosthesis mimics the impedance of an intact portion of the limb, relies on the assumption that the impedance of upper and lower arm change similarly, during energetic interactions with the environment.

\begin{figure}[b]
\centering
\resizebox{2.35in}{!}{\includegraphics{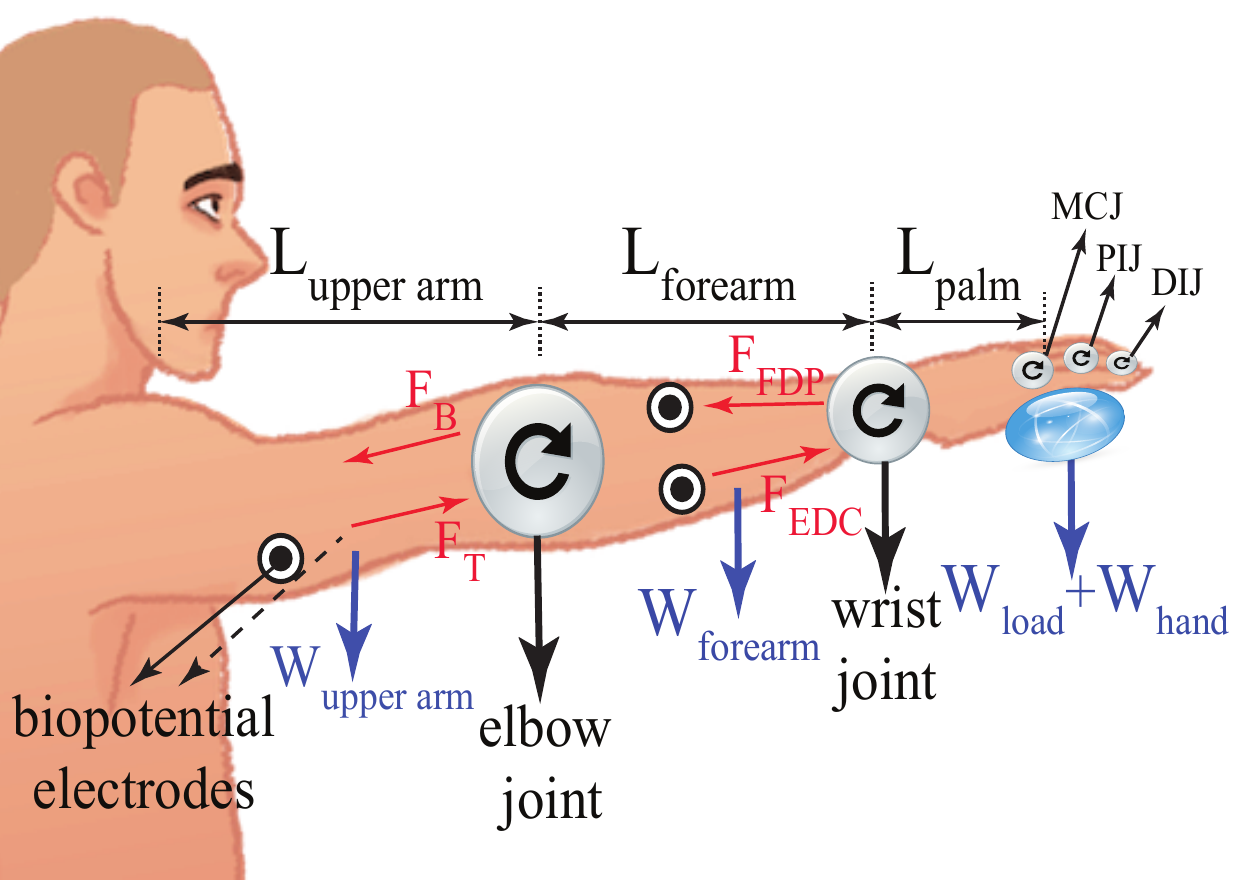}}
\vspace*{-0.25\baselineskip}
\caption{Biomechanical model with the pivots at the wrist and elbow joints, while keeping the arm straight and forward}
\label{biomechanic_models2}
\vspace*{-0.25\baselineskip}
\end{figure}
\begin{figure*}[t]
\centering
\resizebox{5.85in}{!}{\includegraphics{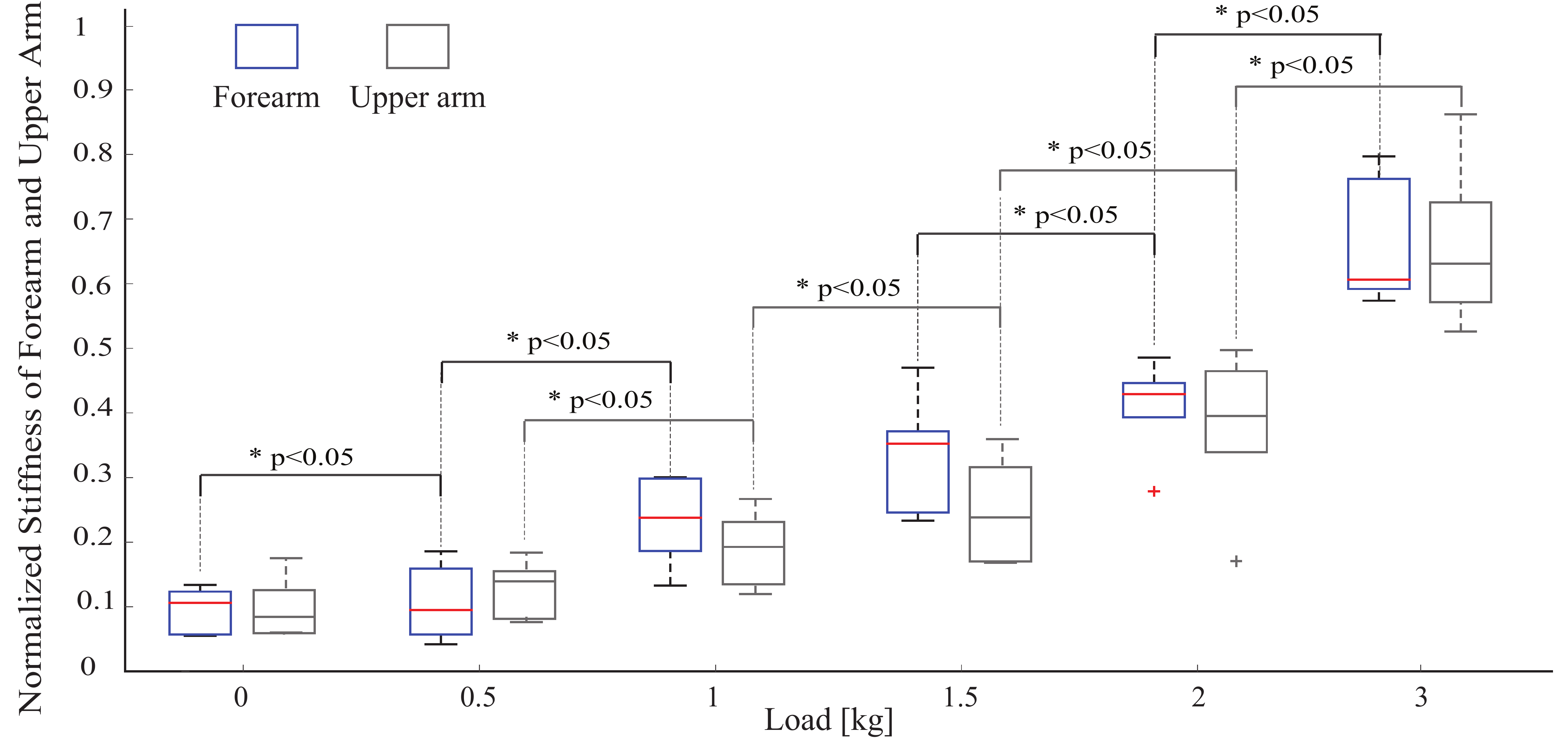}}
\vspace*{-0.5\baselineskip}
\caption{Stiffness estimates from the forearm and the upper arm of participants, while resisting against increasing loads }
\label{impedance_loads}
\vspace*{-.25\baselineskip}
\end{figure*}

We have conducted a series of experiments to test the validity of this assumption. During these experiments, the stiffness of both the forearm and upper arm of participants are estimated through the sEMG signals collected from the relevant antagonistic muscle pairs, using the technique detailed in Section~\ref{Sec:ImpedanceEstimation}.  Eight healthy volunteers took place in the experiments. Before the experiments, all volunteers signed consent forms approved by the IRB of Sabanci University. The experiments were conducted for two tasks:  i) a load bearing task and ii) interaction with the various objects with different impedance characteristics.

The first task aims to observe resistance of the hand, forearm and upper arm against displacement stemmed from the weight of an object with respect to the arm's normal posture. During the first task, participants were asked to keep their arm straight and forward as depicted in Figure~\ref{biomechanic_models2}. The stiffness of upper arm and forearm were estimated as the load at the hand was increased incrementally. In particular, the load was gradually increased from no load to 0.5kg, 1kg, 1.5kg, 2kg, 3kg. Each task was repeated 5 times and each trial lasted about 8 seconds. Sufficient rest time was provided between consecutive trials to prevent muscle fatigue.

Stiffness estimation was performed as detailed in Section~\ref{Sec:ImpedanceEstimation}.  sEMG signals were collected from the antagonistic muscle groups of flexor digitorum profundis and extensor digitorum at the forearm, and  biceps and triceps at the upper arm. Since the stiffness of both the upper arm and forearm were estimated,
two separate biomechanical models were derived around the elbow and wrist joints, respectively. The net torque applied on the joints was calculated considering the weight of the grasped load $W_{load}$, the hand $W_{hand}$, the forearm $W_{forearm}$, and the upper arm $W_{upper arm}$ together with their respective moment arms.

Figure~\ref{impedance_loads} depicts  the estimated stiffness levels at the forearm and the upper arm, under various loading conditions. As expected, as the load is increased, the stiffness of both the upper arm and the forearm increase. As presented in Figure~\ref{impedance_loads}, the change in stiffness levels is statistically significant  between almost all pairs of loading conditions (with $p<0.05$). More importantly, one can observe from these plots that the stiffness increase in the forearm and the upper arm are strongly correlated, and there exist no statistically significant difference between the forearm and the upper arm stiffness levels for each loading condition, for the load bearing task.

The second task tested the adaptation of the upper arm and the forearm impedance levels while interacting with several objects, to mimic common interactions taking place during ADL. In particular, participants started at a rest position, lifted their arm, reached towards the object, grasped it, held it for a while, released it on the table, and returned to their initial configuration. Three different object types were included in the experiment: A sponge, an empty glass, and a water-filled glass were employed for different impedance requirements. Each object was grasped five times and each trial took about 7 seconds. Sufficient rest time was provided to volunteers between sequential trials to prevent muscle fatigue.

The objects were selected such that their manipulation  emphasized different control strategies, ranging from precise motion control to robust force control. Due to the complexity of the task that involved multiple sub-movements, participants' stiffness levels went over continual changes throughout the trials. To quantitatively characterize the correlation between the stiffness of the upper arm and the forearm for each subject, a moving average filter is used to extract average stiffness variations from the instantaneous estimates.  Table~\ref{stiffness_behav_object} presents the Pearson's correlation coefficient for these time series comparisons for each subject. In this table, the concordance correlation coefficients have large values about 0.8, providing strong evidence that the impedance adaptation behaviour of the upper arm and the forearm were in good agreement with each other throughout the complex manipulation task.

\begin{table}[b!]
\caption{Pearson's  correlation coefficient between the stiffness modulation of the upper arm and the forearm muscles}
\vspace*{-\baselineskip}
\label{stiffness_behav_object}
\begin{center}
\begin{tabular}{c || c | c | c} 
    Subject & \hspace{3mm} Sponge \hspace{5mm}& \hspace{4mm} Glass \hspace{4mm} & Water filled glass\\
\hline \hline
    Subject 1 & 0.9246 & 0.9914 & 0.9447 \\
    Subject 2 & 0.9771 & 0.9180	& 0.9000 \\
    Subject 3 & 0.9863 & 0.9134 & 0.9222 \\
    Subject 4 & 0.9000 & 0.9234 & 0.9216 \\
    Subject 5 & 0.9000 & 0.9260 & 0.9715 \\
    Subject 6 & 0.9685 & 0.9158 & 0.9363 \\
    Subject 7 & 0.9062 & 0.9892 & 0.9148 \\
    Subject 8 & 0.9611 & 0.9425 & 0.9775
\end{tabular}
\end{center}
\end{table}

\section{Experimental Evaluation}
\label{Sec:Experimental Evaluation}

We have verified the feasibility and effectiveness of the proposed sEMG based human-machine interface that automatically modulates the impedance of VSA prosthetic hand while users intentionally control the hand position. For this purpose, we have conducted two experiments where the independent control of
hand position and stiffness were demonstrated.

\begin{figure}[b!]
\centering
\resizebox{\columnwidth}{!}{\includegraphics{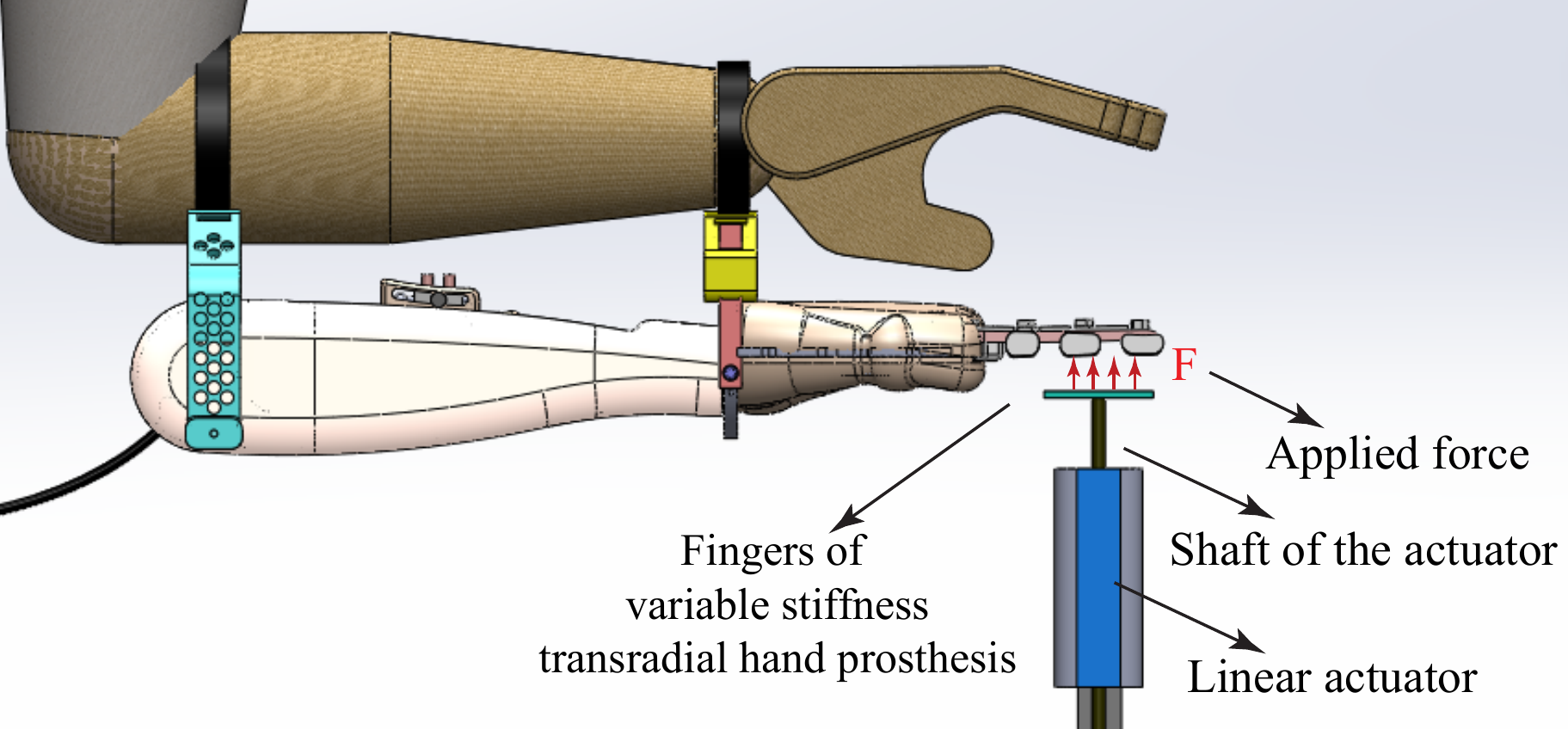}}
\vspace*{-1.5\baselineskip}
\caption{Schematic representation of the experimental setup}
\label{experimentshematic}
\end{figure}

\subsection{Experimental Setup}

Human-subject experiments were conducted using the VSA transradial hand prosthesis detailed in~\cite{Part1}. This prosthesis features underactuated fingers whose position and stiffness can be modulated through antagonistic tendon-based VSA. Throughout the experiments, the transradial hand prosthesis was worn by the volunteers, such that interaction forces with the environment provided direct power coupling with the volunteer. Note that such a feedback is a crucial part of any prosthesis; however, has been neglected in VR based studies~\cite{Okamura2011,okamura2012,okamura2013}. Five healthy volunteers took place in the experiments. The prothesis was worn parallel to the volunteers' lower arm, such that consistent placement of the prosthesis was ensured for proper hand-eye coordination. Before the experiments, all volunteers signed informed consent forms approved by the IRB of Sabanci University.

In the current design, the prosthesis does not feature a thumb, but relies on an passive elastic support that can counteract finger forces. This decision is intentional and helps to keep the system and the controller simple. Our experiences with the volunteers indicate that the passive support is adequate for implementing a wide variety of functional grasps.

Throughout the experiments, sEMG signals were collected from biceps and triceps muscles for stiffness modulation and from trapezius and pectoralis major muscles for position control using a data acquisition system with active electrodes. Stiffness and position references were estimated as discussed in  Section~II and fed to the tracking controller that controlled two geared DC motors under PD control in real-time at 500~Hz through a PC based DAQ card. A direct drive linear actuator combined with a precision position encoder was placed under the fingers of the hand prosthesis as shown in Figure~\ref{experimentshematic} to render forces and measure finger deflections. During the experiments, the gravitational force acting on the linear actuator was compensated with a counter mass, while the linear actuator was force controlled.

\subsection{Experimental Procedure}

Experiments were conducted to test the independent control of the position and the stiffness of the prosthetic hand. These experiments are parallel to the ones presented in~\cite{Part1} to test the VSA transradial hand prosthesis, but differ from them since the sEMG based tele-impedance control interface was used for these experiments, while no such human interface was employed for those experiments. The experiments were repeated here with the sEMG based tele-impedance control interface to verify that the position and the stiffness of the device can be controlled with this interface at a similar performance level as an externally modulated controller.

The experiments were composed of two tasks with 10 repetitions for each condition of each task. During the first task, the position of the VSA hand prosthesis was kept constant at $0^{\circ}$ while the stiffness of VSA was adjusted by the volunteers to five distinct stiffness values that correspond to a low, three intermediate, and a high stiffness level for the fingers. The stiffness of the fingers were experimentally determined by applying a linearly increasing force to flex the fingers and recording their deflection.


During the second task, the stiffness of the VSA hand prosthesis was kept constant at its intermediate level by the volunteers, while the position of the VSA was adjusted by the volunteers to three distinct position values that correspond to low, intermediate, and high flexion of the fingers. The position of the fingers was determined by recording the position of the linear actuator under zero force control, while the stiffness of the fingers were determined by applying a constant force to resist the flexion of the fingers at their equilibrium position and recording the resulting deflection.

\subsection{Experimental Results}

\begin{figure}[t!]
\centering
\resizebox{\columnwidth}{!}{\includegraphics{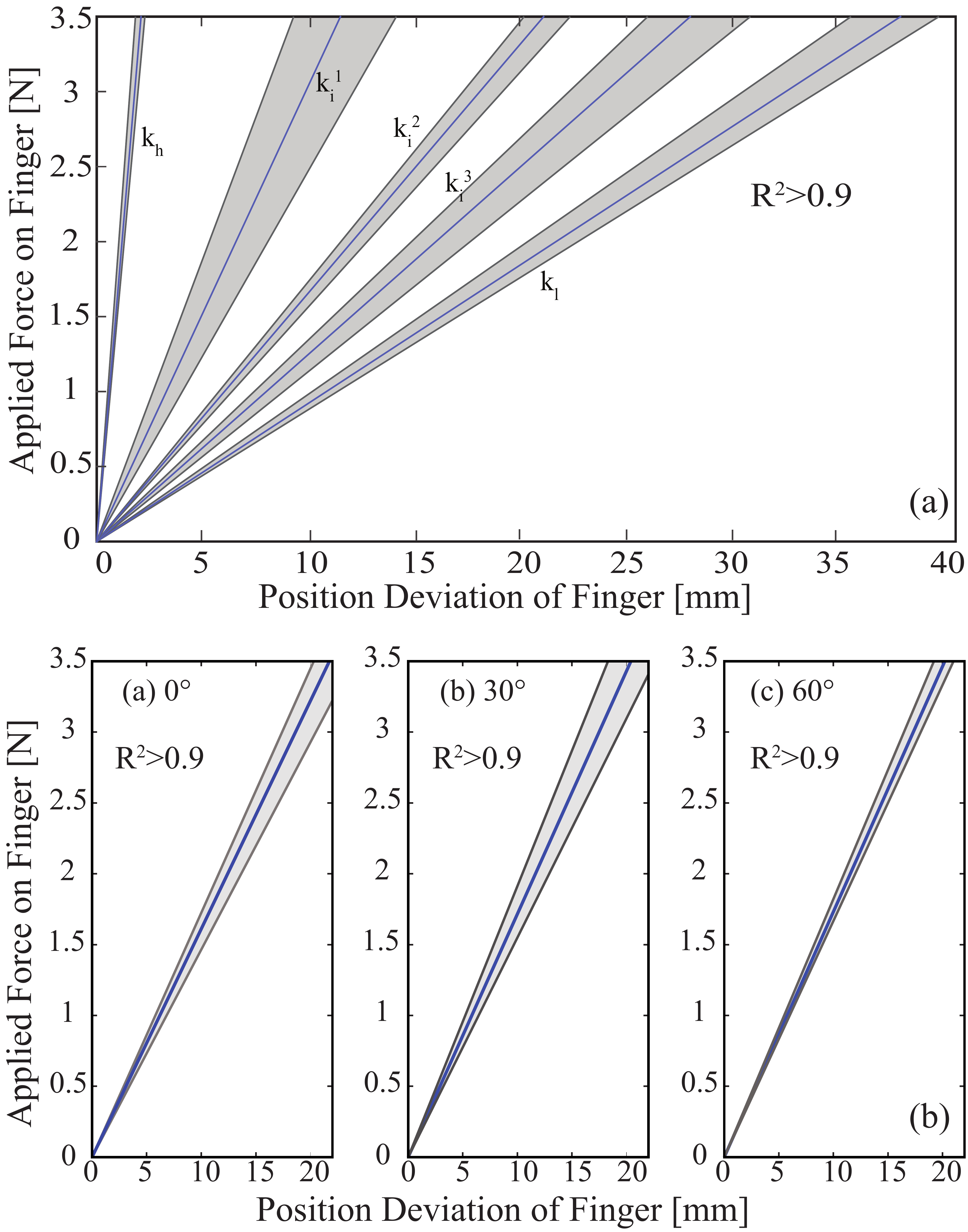}}
\vspace*{-1.2\baselineskip}
\caption{(a) Stiffness modulation of hand prosthesis through sEMG based tele-impedance control. (b) Position control of hand prosthesis through sEMG based tele-impedance control. In the figures, gray zones present the results of each trial and the blue lines represent the average value of ten trials.}
\label{results_emg_all}
\end{figure}

Figure~\ref{results_emg_all}(a) presents the experimental results for the case when the volunteers adjusted the VSA stiffness to five distinct values that correspond to a low, three intermediate and a high stiffness level for the fingers, while the finger positions were kept constant. In particular, shaded regions represent all the linear fits recorded for 10 trials, while the dark line represents their mean. The slopes of these lines indicate that the high, three intermediate and the low stiffness for the fingers were $k_{h}$=1.7~N/mm, $k_{i}^{1}$=0.3~N/mm, $k_{i}^{2}$=0.16~N/mm, $k_{i}^{3}$=0.12~N/mm and $k_{l}$=0.091~N/mm, respectively. The $R^2$ values for these linear fits are higher than~0.97.

Figure~\ref{results_emg_all}(b) presents the experimental results for the case when the volunteers kept the VSA stiffness at an intermediate level, while the finger positions of the fingers were regulated by the volunteers to $0^{\circ}$, $30^{\circ}$, and $60^{\circ}$, respectively. Once again, the shaded regions represent all the linear fits recorded for 10 trials, while the dark line represents their mean. The slopes of these lines indicate that stiffness level of the fingers were
$k_{0^\circ}$=0.16~N/mm,  $k_{30^\circ}$=0.17~N/mm, and $k_{60^\circ}$=0.17~N/mm, respectively. The $R^2$ values for these linear fits are higher than~0.98.

The fingers' response shown in Figure~\ref{results_emg_all} closely matches the characteristics of human fingers, as presented in~\cite{Howe1985}. The  characterization results are also compatible with the results presented in~\cite{finger_extension}, as flexion/extension movements performed by an anatomically human-like robotic index finger necessitate similar amount of muscle forces.

\begin{figure*}[t]
\centering
\resizebox{2.05\columnwidth}{!}{\includegraphics{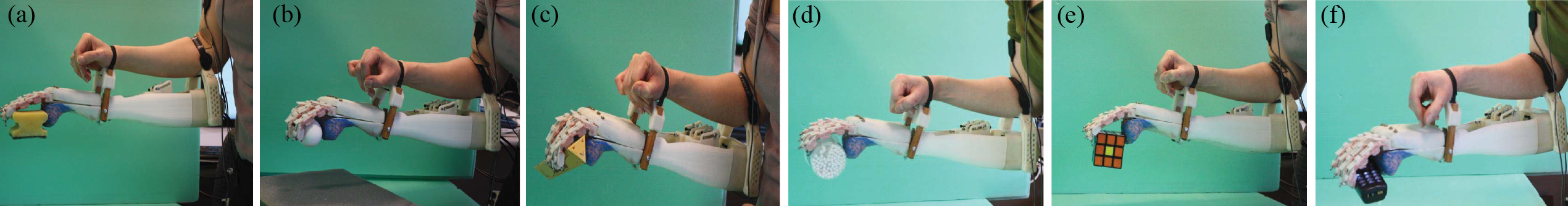}}
\vspace*{-1\baselineskip}
\caption{Demonstration of variable stiffness transradial hand prosthesis performing various grasps with the sEMG based tele-impedance control interface, while interacting with (a) a deformable object, (b) a fragile object, (c) a triangular rigid object, (d) a cylindrical rigid object, (f) a square rigid object, and (g) a rectangular rigid object.}
\vspace*{0\baselineskip}
\label{illustrative}
\end{figure*}

Experimental results indicate that the sEMG based impedance controlled VSA hand prosthesis possesses very similar performance to the case with external reference generator, as presented in~\cite{Part1}. In particular, volunteers were able to modulate their stiffness levels to minimum and maximum stiffness limits of the prosthetic hand, as well as to various intermediate ranges, by means of the sEMG based tele-impedance control. These results provide evidence that the stiffness and position of the transradial hand prosthesis can be controlled independently by users, with high repeatability. 

\section{Illustrative Experiments and Discussions}

Given that only the position and the stiffness of the drive tendon can be directly regulated by the volunteers, in general, the resulting position and the stiffness of the fingers  depend on the interaction. To test the usefulness of the sEMG based tele-impedance control interface of the variable stiffness transradial hand prosthesis, the device was attached to several volunteers, as shown in Figure~\ref{experimentshematic} and the volunteers were given control of the position and stiffness of the prosthesis through the sEMG based tele-impedance controller. In particular, sEMG signals measured from the surface of the upper arm were used to automatically adjust the stiffness level of the prosthesis to that of the upper arm, while the position regulation was intentionally controlled by the volunteers by moving their shoulder muscles. With this natural control interface, volunteers were asked to grasp a wide variety of objects with different shapes (e.g., cylindrical, rectangular, elliptic or unstructured) and compliance (e.g., rigid, soft, elastic).

The volunteers were successful at grasping a wide variety of objects as shown in Figure~\ref{illustrative}. In Figure~\ref{illustrative}(a) a deformable sponge, in Figure~\ref{illustrative}(b), a fragile raw egg were grasped by the volunteers with low stiffness to prevent damage to the objects. In Figures~\ref{illustrative}(c)--(f), rigid objects with various shapes were grasped by the volunteers using different stiffness levels. Videos demonstrating several illustrative grasps by a volunteer are available at \url{https://youtu.be/fGFIKSSmtDg}.

The proposed tele-impedance controller interface emphasizes simplicity, ease of use and adaptability; hence, implements automatic modulation of prosthetic hand stiffness to match that of upper arm, while intentional control of position of the underactuated prosthetic hand is left to the user. Under the observation that humans tend to modulate the impedance of their limb as a whole  while executing different tasks (as shown in Section~VII), the tele-impedance controller implemented for the prosthesis automatically modulates the stiffness of the hand to match that of the intact part of the arm. Automatic stiffness modulation increases the dexterity of the prosthetic hand, without introducing complexity to the human control interface.

Successful interactions with the prosthesis depends on the amputee making proper decisions on how to interact with the object under visual feedback and physical coupling. Our extensive experiments with healthy volunteers indicate that humans are very skillful at learning how to interact with the environment with such a device under the proposed sEMG based control interface. All volunteers were able to adapt to the device with a few minutes and successfully complete the required manipulation tasks without any prior training. Furthermore, it has been observed that stiffness modulation property is effective in increasing the performance of the transradial prosthesis.

Volunteers suffer from the high complexity of the controller when intentional control of both the stiffness and the position of the device is left to the user.  During our tests, volunteers indicated a strong preference of the automatic impedance adjustment property. Furthermore, it has been observed that volunteers are more successful at interactions when the impedance of the prosthesis is automatically adjusted.

\section{Conclusion and Future Work}

Tele-impedance control of a VSA prosthetic hand is implemented through stiffness and position estimates decoded from sEMG signals of muscle groups embedded in the upper arm, chest, and shoulder. In particular, IMCJ method is used to estimate stiffness of intact upper arm through agonist/antagonist muscle pairs, while shoulder/chest muscles are employed to estimate position references. Then, these stiffness and position estimates are used to control a VSA prosthetic hand.

The feasibility of tele-impedance control through the proposed human machine interface is demonstrated with two human subject experiments, where the position and the stiffness of the VSA prosthetic hand were successfully modulated.  The results demonstrate that both position and stiffness  estimations from sEMG signals are adequate for control of a VSA transradial hand prosthesis.

VSA hand prosthesis together with the proposed control interface necessitates less effort and  concentration to control, and is easier for the amputee to learn to use.  Impedance modulation takes place naturally from task to task or while performing a task, i.e. activities of daily living, without requiring amputees' attention, and this feature improves the performance of the prosthesis while interacting with unstructured environments.

Future work includes further testing with amputees. Furthermore, integration of additional feedback pathways, such as vibrotactile feedback into the system in addition to the currently available visual feedback can be pursued. Another line of future research is the online determination of sEMG related coefficients to avoid prior tests with the amputee.

\section*{Acknowledgment}
This work has been partially supported by Tubitak Grants 107M337, 109M020, 111M186, 115M698, Marie Curie IRG Rehab-DUET and Sabanci University IRP.

\ifCLASSOPTIONcaptionsoff
  \newpage
\fi

\bibliographystyle{IEEEtran}
\bibliography{mymanuscript}

\begin{thebibliography}{10}
\providecommand{\url}[1]{#1}
\csname url@samestyle\endcsname
\providecommand{\newblock}{\relax}
\providecommand{\bibinfo}[2]{#2}
\providecommand{\BIBentrySTDinterwordspacing}{\spaceskip=0pt\relax}
\providecommand{\BIBentryALTinterwordstretchfactor}{4}
\providecommand{\BIBentryALTinterwordspacing}{\spaceskip=\fontdimen2\font plus
\BIBentryALTinterwordstretchfactor\fontdimen3\font minus
  \fontdimen4\font\relax}
\providecommand{\BIBforeignlanguage}[2]{{%
\expandafter\ifx\csname l@#1\endcsname\relax
\typeout{** WARNING: IEEEtran.bst: No hyphenation pattern has been}%
\typeout{** loaded for the language `#1'. Using the pattern for}%
\typeout{** the default language instead.}%
\else
\language=\csname l@#1\endcsname
\fi
#2}}
\providecommand{\BIBdecl}{\relax}
\BIBdecl

\bibitem{Marino15}
M.~Marino, S.~Pattni, M.~Greenberg, A.~Miller, E.~Hocker, S.~Ritter, and
  K.~Mehta, ``Access to prosthetic devices in developing countries: Pathways
  and challenges,'' in \emph{{IEEE} Global Humanitarian Technology Conference
  (GHTC)}, 2015, pp. 45--51.

\bibitem{Graham2007}
K.~Ziegler-Graham, E.~MacKenzie, P.~Ephraim, T.~Travison, and R.~Brookmeyer,
  ``Estimating the prevalence of limb loss in the united states: 2005 to
  2050,'' \emph{Archives of Physical Medicine and Rehabilitation}, vol.~89,
  no.~3, pp. 422--429, 2007.

\bibitem{Lusardi13}
M.~Lusardi, M.~Jorge, and C.~Nielsen, \emph{Orthotics and Prosthetics in
  Rehabilitation}, 3rd~ed.\hskip 1em plus 0.5em minus 0.4em\relax Saunders
  Elsevier, 2013.

\bibitem{Millstein}
S.~Millstein, D.~Bain, and G.~Hunter, ``A review of employment patterns of
  industrial amputees-{F}actors influencing rehabilitation,'' \emph{Prosthet
  Orthot Int}, vol.~9, no.~2, pp. 69--78, 1985.

\bibitem{Biddiss}
E.~Biddiss and T.~Chau, ``Upper limb prosthesis use and abandonment: a survey
  of the last 25 years,'' \emph{Prosthet Orthot Int.}, vol.~31, no.~3, pp.
  236--257, Sep 2007.

\bibitem{McFarland}
L.~McFarland, S.~Hubbard~Winkler, W.~A., J.~M., and A.~Esquenazi, ``Unilateral
  upper-limb loss: Satisfaction and prosthetic-device use in veterans and
  servicemembers from vietnam and oif/oef conflicts,'' \emph{J Rehabil Res
  Dev.}, vol.~47, no.~4, pp. 299--316, 2010.

\bibitem{Wang}
N.~Wang, K.~Lao, and X.~Zhang, ``Design and myoelectric control of an
  anthropomorphic prosthetic hand,'' \emph{Journal of Bionic Engineering},
  vol.~14, no.~1, pp. 47--59, 2017.

\bibitem{Naik}
G.~R. Naik, A.~H. Al-Timemy, and H.~T. Nguyen, ``Transradial amputee gesture
  classification using an optimal number of semg sensors: An approach using ica
  clustering,'' \emph{IEEE Transactions on Neural Systems and Rehabilitation
  Engineering}, vol.~24, no.~8, pp. 837--846, 2016.

\bibitem{Riillo}
F.~Riillo, L.~Quitadamo, F.~Cavrini, E.~Gruppioni, C.~Pinto, N.~C. Pasto,
  L.~Sbernini, L.~Albero, and G.~Saggio, ``Optimization of emg-based hand
  gesture recognition: Supervised vs. unsupervised data preprocessing on
  healthy subjects and transradial amputees,'' \emph{Biomedical Signal
  Processing and Control}, vol.~14, pp. 117--125, 2014.

\bibitem{ilimb}
\BIBentryALTinterwordspacing
{Touch Bionics Inc.}, 2019. [Online]. Available:
  \url{https://www.ossur.com/prosthetic-solutions/products/touch-solutions/i-limb-ultra}
\BIBentrySTDinterwordspacing

\bibitem{Ottobockhand}
\BIBentryALTinterwordspacing
Ottobock, 2019. [Online]. Available:
  \url{http://www.ottobock.com/cps/rde/xchg/ob_com_en/hs.xsl/384.html}
\BIBentrySTDinterwordspacing

\bibitem{bebionic}
\BIBentryALTinterwordspacing
bebionic, 2019. [Online]. Available:
  \url{https://www.ottobockus.com/prosthetics/upper-limb-prosthetics/solution-overview/bebionic-hand/}
\BIBentrySTDinterwordspacing

\bibitem{mchand}
\BIBentryALTinterwordspacing
{Motion Control Inc.}, 2019. [Online]. Available:
  \url{http://www.utaharm.com/motion-control-electric-hand-terminal-device.php}
\BIBentrySTDinterwordspacing

\bibitem{Kurzynski}
M.~Kurzynski, M.~Krysmann, P.~Trajdos, and A.~Wolczowski, ``Multiclassifier
  system with hybrid learning applied to the control of bioprosthetic hand,''
  \emph{Computers in Biology and Medicine}, vol.~69, pp. 286--297, 2016.

\bibitem{Guo}
W.~Guo, X.~Sheng, H.~Liu, and X.~Zhu, ``Toward an enhanced human-machine
  interface for upper limb prosthesis control with combined {EMG} and {NIRS}
  signals,'' \emph{{IEEE} Transactions on Human-Machine Systems}, no.~99, pp.
  1--12, 2017.

\bibitem{Kyranou}
I.~Kyranou, A.~Krasoulis, M.~S. Erden, K.~Nazarpour, and S.~Vijayakumar,
  ``Real-time classification of multi-modal sensory data for prosthetic hand
  control,'' in \emph{{IEEE} International Conference on Biomedical Robotics
  and Biomechatronics}, 2016, pp. 536--541.

\bibitem{Herle}
S.~Herle, ``Design of a reference signal generator for an upper limb prosthesis
  myoelectric controller,'' in \emph{{IEEE} International Conference on
  Automation, Quality and Testing, Robotics (AQTR)}, May 2016, pp. 1--6.

\bibitem{Atkins}
D.~Atkins, D.~Heard, and W.~Donovan, ``Epidemiologic overview of individuals
  with upper limb loss and their reported research priorities,'' \emph{Journal
  of Prosthetics and Orthotics}, vol.~8, no.~1, pp. 2--11, 1996.

\bibitem{Perreault}
E.~J. Perreault, R.~Kirsch, and P.~Crago, ``Multijoint dynamics and postural
  stability of the human arm,'' \emph{Experimental Brain Research}, vol. 157,
  no.~4, pp. 507--517, 2004.

\bibitem{Popescu}
F.~Popescu, J.~Hidler, and W.~Rymer, ``Elbow impedance during goal-directed
  movements,'' \emph{Experimental Brain Research}, vol. 152, no.~1, pp. 17--28,
  2003.

\bibitem{Franklin2003}
D.~Franklin, E.~Burdet, R.~Osu, M.~Kawato, and T.~Milner, ``Functional
  significance of stiffness in adaptation of multijoint arm movements to stable
  and unstable dynamics,'' \emph{Experimental Brain Research}, vol. 151, no.~2,
  pp. 145--157, 2003.

\bibitem{Milner2003}
D.~Franklin and T.~Milner, ``Adaptive control of stiffness to stabilize hand
  position with large loads,'' \emph{Experimental Brain Research}, vol. 152,
  no.~2, pp. 211--220, 2003.

\bibitem{Franklin2004}
D.~Franklin, U.~So, M.~Kawato, and T.~Milner, ``Impedance control balances
  stability with metabolically costly muscle activation,'' \emph{Journal of
  Neurophysiology}, vol.~92, no.~5, pp. 3097--3105, 2004.

\bibitem{Kawato2003}
D.~Franklin, M.~Kawato, U.~So, and T.~Milner, ``Interacting with our
  environment: Impedance control balances stability and metabolic cost,'' in
  \emph{{IEEE} Engineering in Medicine and Biology Society}, vol.~2, Sept 2003,
  pp. 1440--1443.

\bibitem{hogan2002}
N.~Hogan, ``Skeletal muscle impedance in the control of motor actions,''
  \emph{Journal of Mechanics in Medicine and Biology}, vol.~02, no. 03n04, pp.
  359--373, 2002.

\bibitem{hoganpart1}
C.~Abul-Haj and N.~Hogan, ``Functional assessment of control systems for
  cybernetic elbow prostheses. i. description of the technique.'' \emph{{IEEE}
  Trans. Biomed. Eng.}, vol.~37, no.~11, pp. 1025--1036, 1990.

\bibitem{hoganpart2}
------, ``Functional assessment of control systems for cybernetic elbow
  prostheses. ii. application of the technique.'' \emph{{IEEE} Trans. Biomed.
  Eng.}, vol.~37, no.~11, pp. 1037--1047, 1990.

\bibitem{twente}
S.~Rao, R.~Carloni, and S.~Stramigioli, ``Stiffness and position control of a
  prosthetic wrist by means of an {EMG} interface,'' \emph{IEEE EMBS}, pp.
  495--498, 2010.

\bibitem{Okamura2011}
A.~Blank, A.~Okamura, and L.~Whitcomb, ``Task-dependent impedance improves user
  performance with a virtual prosthetic arm,'' in \emph{{IEEE} International
  Conference on Robotics and Automation (ICRA)}, May 2011, pp. 2235--2242.

\bibitem{okamura2012}
------, ``User comprehension of task performance with varying impedance in a
  virtual prosthetic arm: A pilot study,'' \emph{4th {IEEE RAS EMBS}
  International Conference on Biomedical Robotics and Biomechatronics}, pp.
  500--507, 2012.

\bibitem{okamura2013}
------, ``Task-dependent impedance and implications for upper-limb prosthesis
  control,'' \emph{International Journal of Robotics Research}, vol.~3, no.~6,
  pp. 827--846, 2013.

\bibitem{hocaoglue12}
E.~Hocaoglu and V.~Patoglu, ``Tele-impedance control of a variable stiffness
  prosthetic hand,'' in \emph{IEEE International Conference on Robotics and
  Biomimetics (ROBIO)}, 2012, pp. 1576--1582.

\bibitem{Part2}
------, ``{sEMG}-based natural control interface for a variable stiffness
  transradial hand prosthesis,'' \emph{{IEEE/ASME} Transactions on
  Mechatronics}, 2019, (under review).

\bibitem{myohand1}
S.~A. Dalley, T.~E. Wiste, T.~J. Withrow, and M.~Goldfarb, ``Design of a
  multifunctional anthropomorphic prosthetic hand with extrinsic actuation,''
  \emph{{IEEE/ASME} Transactions on Mechatronics}, vol.~14, no.~6, pp.
  699--706, Dec 2009.

\bibitem{myohand2}
T.~Lenzi, J.~Lipsey, and J.~W. Sensinger, ``The {RIC} arm \textendash~a small
  anthropomorphic transhumeral prosthesis,'' \emph{{IEEE/ASME} Transactions on
  Mechatronics}, vol.~21, no.~6, pp. 2660--2671, Dec 2016.

\bibitem{myohand3}
D.~A. Bennett, J.~E. Mitchell, D.~Truex, and M.~Goldfarb, ``Design of a
  myoelectric transhumeral prosthesis,'' \emph{{IEEE/ASME} Transactions on
  Mechatronics}, vol.~21, no.~4, pp. 1868--1879, Aug 2016.

\bibitem{Kim2019}
N.~{Kim}, S.~{Yun}, and D.~{Shin}, ``A bio-inspired lightweight wrist for
  high-dof robotic prosthetic arms,'' \emph{IEEE/ASME Transactions on
  Mechatronics}, pp. 1--1, 2019.

\bibitem{Part1}
E.~Hocaoglu and V.~Patoglu, ``Design, implementation and evaluation of a
  variable stiffness transradial hand prosthesis,'' \emph{{IEEE/ASME}
  Transactions on Mechatronics}, 2019, (under review).

\bibitem{Gomi05041996}
H.~Gomi and M.~Kawato, ``Equilibrium-point control hypothesis examined by
  measured arm stiffness during multijoint movement,'' \emph{Science}, vol.
  272, no. 5258, pp. 117--120, 1996.

\bibitem{Burdet2000}
E.~Burdet, R.~Osu, D.~W. Franklin, T.~Yoshioka, T.~E. Milner, and M.~Kawato,
  ``A method for measuring endpoint stiffness during multi-joint arm
  movements,'' \emph{Journal of Biomechanics}, vol.~33, no.~12, pp. 1705--1709,
  2000.

\bibitem{Burdet2001}
E.~Burdet, R.~Osu, D.~W. Franklin, T.~E. Milner, and M.~Kawato, ``The central
  nervous system stabilizes unstable dynamics by learning optimal impedance,''
  \emph{Nature 414}, pp. 446--449, June 2001.

\bibitem{shortandlong}
R.~Osu, D.~W. Franklin, H.~Kato, H.~Gomi, K.~Domen, T.~Yoshioka, and M.~Kawato,
  ``Short- and long-term changes in joint co-contraction associated with motor
  learning as revealed from surface {EMG},'' \emph{J Neurophysiol}, vol.~88,
  pp. 991--1004, 2002.

\bibitem{Gomi1998}
H.~Gomi and R.~Osu, ``Task-dependent viscoelasticity of human multijoint arm
  and its spatial characteristics for interaction with environments,''
  \emph{Journal of Neuroscience}, vol.~18, pp. 8965--8978, Nov. 1998.

\bibitem{Hunter_Kearney_1982}
I.~W. Hunter and R.~E. Kearney, ``Dynamics of human ankle stiffness: variation
  with mean ankle torque,'' \emph{Journal of Biomechanics}, vol.~15, no.~10,
  pp. 747--752, 1982.

\bibitem{Basmajian}
J.~V. Basmajian and C.~J. {De Luca}, \emph{Muscles Alive}.\hskip 1em plus 0.5em
  minus 0.4em\relax Baltimore, MD: Williams and Wilkins, 1985.

\bibitem{Kuechle1997}
D.~Kuechle, S.~R. Newman, E.~Itoi, B.~Morrey, and K.~N. An, ``Shoulder muscle
  moment arms during horizontal flexion and elevation,'' \emph{Journal of
  Shoulder and Elbow Surgery}, vol.~6, no.~5, pp. 429--439, 1997.

\bibitem{Murray1995}
W.~Murray, S.~Delp, and T.~Buchanan, ``Variation of muscle moment arms with
  elbow and forearm position,'' \emph{J Biomech}, vol.~28, no.~5, pp. 513--525,
  1995.

\bibitem{mus_ftg4}
M.~Al-Mulla, F.~Sepulveda, and M.~Colley, ``A review of non-invasive techniques
  to detect and predict localised muscle fatigue,'' \emph{Sensors}, vol.~11,
  no.~4, pp. 3545--94, Mar 2011.

\bibitem{mus_ftg2}
R.~Merletti, A.~Rainoldi, and D.~Farina, \emph{Myoelectric Manifestations of
  Muscle Fatigue}.\hskip 1em plus 0.5em minus 0.4em\relax John Wiley $\&$ Sons,
  Inc. Hoboken, New Jersey, 2005, pp. 233--258.

\bibitem{deluca84}
C.~De~Luca, ``Myoelectrical manifestations of localized muscular fatigue in
  humans,'' Crit Rev. Biomed. Eng., pp. 251--279, Feb 1984.

\bibitem{mus_ftg1}
M.~Cifrek, V.~Medved, S.~Tonkovic, and S.~Ostojic, ``Surface {EMG} based muscle
  fatigue evaluation in biomechanics,'' \emph{Clinical Biomechanics}, vol.~24,
  no.~4, pp. 327--340, Mar 2009.

\bibitem{mus_ftg3}
M.~Bilodeau, S.~Schindler-Ivens, D.~Williams, R.~Chandran, and S.~Sharma,
  ``{EMG} frequency content changes with increasing force and during fatigue in
  the quadriceps femoris muscle of men and women,'' \emph{Journal of
  Electromyography and Kinesiology}, vol.~13, no.~1, pp. 83--92, Feb 2003.

\bibitem{Migliore2007}
S.~A. Migliore, E.~A. Brown, and S.~P. DeWeerth, ``Novel nonlinear elastic
  actuators for passively controlling robotic joint compliance,'' \emph{Journal
  of Mechanical Design}, vol. 129, no.~4, pp. 406--412, 2007.

\bibitem{EnglishRussell}
C.~English and D.~Russell, ``Mechanics and stiffness limitations of a variable
  stiffness actuator for use in prosthetic limbs,'' \emph{Mechanism and Machine
  Theory}, vol.~34, no.~1, pp. 7--25, Jan. 1999.

\bibitem{Howe1985}
A.~Howe, D.~Thompson, and V.~Wright, ``Reference values for metacarpophalangeal
  joint stiffness in normals,'' \emph{Annals of the Rheumatic Diseases},
  vol.~44, pp. 469--476, July 1985.

\bibitem{finger_extension}
Y.~Matsuoka and P.~Afshar, ``Neuromuscular strategies for dynamic finger
  movements: a robotic approach,'' in \emph{International Conference of the
  IEEE Engineering in Medicine and Biology Society}, vol.~2, Sept 2004, pp.
  4649--4652.

\end{thebibliography}


%
%

\begin{IEEEbiography}[{\includegraphics[width=1in,height=1.25in,clip,keepaspectratio]{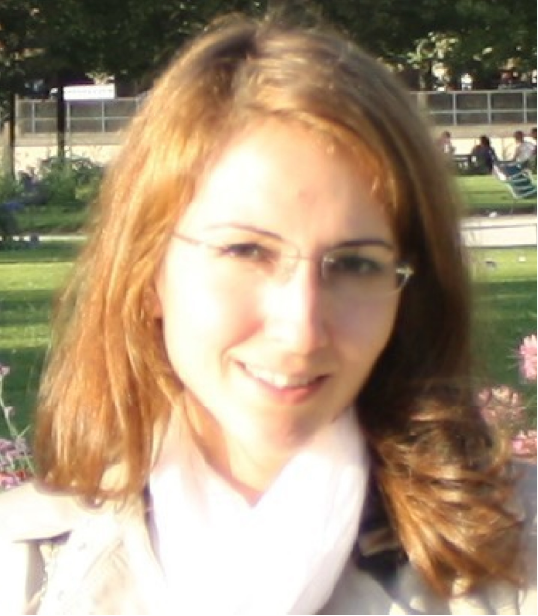}}]
{Elif Hocaoglu} received her Ph.D. degree in Mechatronics Engineering from the Sabanci University, Istanbul in
2014. She worked as a post doctoral researcher at Advanced Robotics Laboratory at Italian Institute of Technology and then Human Robotics Laboratory at Imperial College London. Currently, she is an assistant professor at Istanbul Medipol University. Her research is in the area of physical human-machine interaction, in particular sEMG based control interfaces with applications to upper extremity prostheses, rehabilitation and assistive devices. Her research extends to wearable robotics.
\end{IEEEbiography}

\begin{IEEEbiography}[{\includegraphics[width=1in,height=1.25in,clip,keepaspectratio]{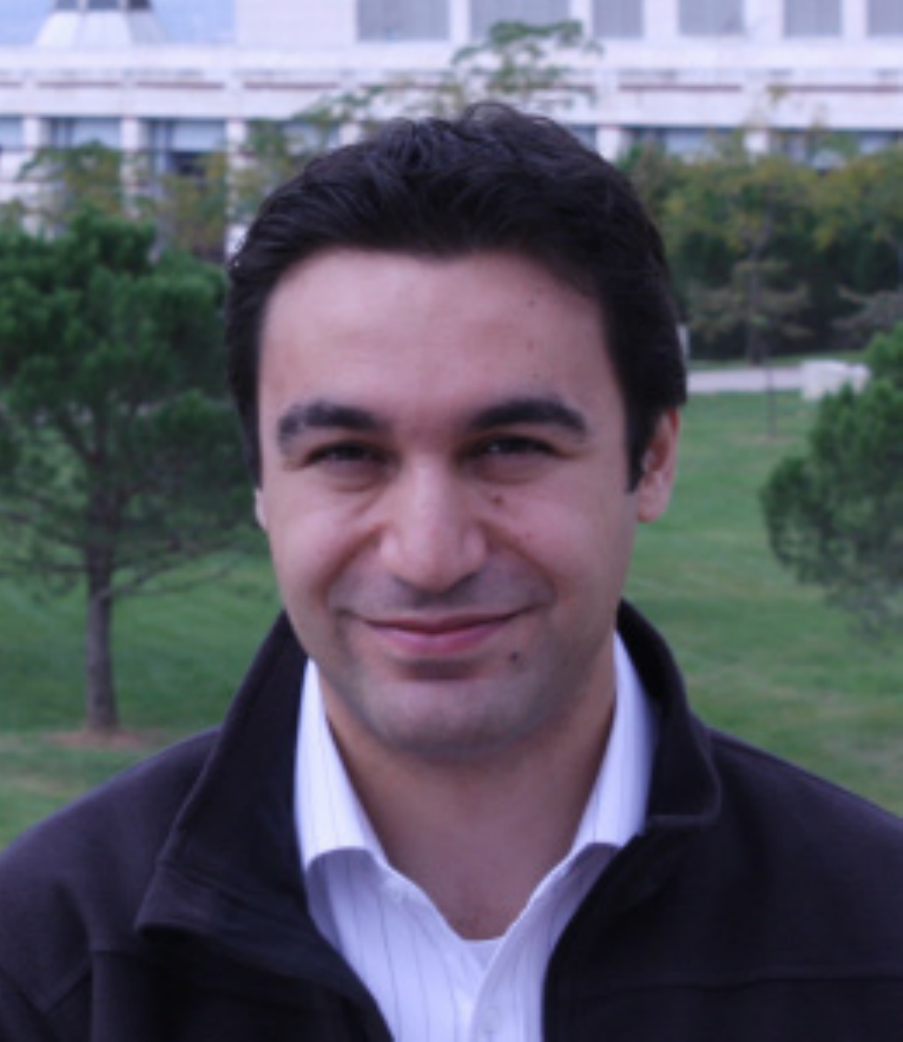}}]
{Volkan Patoglu} received his Ph.D. degree in
Mechanical Engineering from the University of Michigan, Ann Arbor in
2005. He worked as a post doctoral research associate in
Mechatronics and Haptic Interfaces Laboratory at Rice University.
Currently, he is a professor at Sabanci University. His
research is in the area of physical human-machine interaction, in
particular, design and control of force feedback robotic systems
with applications to rehabilitation and skill training. His research
extends to cognitive robotics.

Dr. Patoglu has served as an associate editor for the IEEE Transactions on Haptics (2013--2017)
and is currently an associate editor for the IEEE Transactions on Neural Systems and Rehabilitation Engineering and IEEE Robotics and Automation Letters.
\end{IEEEbiography}

\vfill

\end{document}